\documentclass[%
  reprint,
superscriptaddress,
 amsmath,amssymb,
 aps,
prl,
nofootinbib
]{revtex4-2}

\usepackage{amsmath,amssymb,amsthm,mathrsfs,amsfonts,dsfont,amstext}
\usepackage{graphicx}
\usepackage{dcolumn}
\usepackage{bm}
\usepackage{hyperref}
\usepackage[dvipsnames]{xcolor}
\usepackage{orcidlink}
\usepackage{multirow}
\usepackage{aas_macros}
\usepackage{slashed}
\usepackage{subfigure}
\usepackage{physics}

\hypersetup{
  colorlinks   = true, 
  urlcolor     = Maroon, 
  linkcolor    = PineGreen, 
  citecolor   = Maroon 
}

\makeatletter
\def\maketitle{
\@author@finish
\title@column\titleblock@produce
\suppressfloats[t]}
\makeatother

\begin{document}


\title{Emergent Theory for Constitutive Relations in Classical Defect Systems}

\author{Hongrui Xu}
\email{21210190019@m.fudan.edu.cn}
\affiliation{Department of Physics, State Key Laboratory of Surface Physics, and Key Laboratory of Micro and Nano Photonic Structures (MOE),\\
	Fudan University, Shanghai 200438, China}

\begin{abstract}
Optical, acoustic, hydrodynamic, and thermal defect systems are often studied by analogy with each other. This may indicate that we may find a emergent theory for constitutive relations of classical defect systems. Start with thermal systems, we put up with a bootstrap method to describe classical transport. We conjecture that Landau-Khalatnikov equation could provide heat constitutive relations when taking heat flux as order parameter. We show that the corresponding effctive Lagrangian has similar form of that of Wick-rotated complex scalar field theory at non-relativistic limit, and only in perfect conducting situation that this system is a canonical ensemble. We argue that our method could be generalized to other systems besides thermal ones. By analogy with ferroelectrics, we propose the thermal domain model, which is a conserved current XY model. Phase transition of the thermal system is qualitatively discussed in this model.
\end{abstract}

\maketitle

\newpage


Defect systems exhibit novel physical properties through artificial manipulations of defects. Throughout the history of defect system research, 2-dimensional  \cite{cao2018unconventional,yankowitz2019tuning}, Optical \cite{veselago1967electrodynamics,shen2024nondiffracting,tang2021modeling}, acoustic \cite{ZYLiu,zhu2023topological,qi2020acoustic,han2025observation,torrent2008acoustic}, hydrodynamic \cite{xu2024hydrodynamic,morton2008hydrodynamic,oh2012design} and thermal \cite{yue2021thermal,fan2008shaped,xu2020controlling,ren2010berry}
defect materials
are often studied by analogy with each other. An intriguing question is whether this phenomenon implies some underlying physical mechanism. Can we establish a unified theory to describe classical transport in all types of defect systems? We try to handle with this problem by studying constitutive relations.

First we start with thermal constitutive relations. Inspired by the Landau paradigm in ferroelectric materials \cite{landauKhalatnikov,LK2021,Size,liu2025phase,zhang2021electrocaloric}, we treat heat flux as order parameter. We hypothesize that classical heat constitutive relations can be derived from Landau-Khalatnikov equation. We find that heat conductivity can be interpreted as the inverse of the Landau coefficient times a coupling constant. This effective theory has similar form as complex scalar field at non-relativistic limit with imaginary time \cite{Amateur, tong2016lectures}. The discussion is presented in Supplemental Material. We argue that this method is not only valid for thermal systems, but for other systems as well.

In analogy to ferroelectric materials, we introduce the thermal domains in thermal systems \cite{grunebohm2023influence,ye2018giant,yu2024large}.  We provide quantitative description of first order phase transition using thermal domain model. The formation of latent heat is illustrated in our model.

{\it{Heat constitutive relations}}\quad
Heat constitutive relations have long been regarded as phenomenological equations. The first discovered heat constitutive relation, Fourier's law, was recognized to violate the principle of causality due to its instantaneous propagation of thermal signals. This theoretical limitation prompted the development of the Maxwell-Cattaneo equation. Since then, numerous modified versions of heat constitutive relations have been systematically developed and studied \cite{zhmakin2023non}. Within hydrodynamic frameworks, the Fourier heat constitutive relation appears naturally as the leading-order term in a derivative expansion of the U(1) conserved current \cite{Kailidis:2023ntq,jain2023S-K,HongLiuLecture,Kapustin_2023}. However, non-Fourier heat constitutive relations have rarely been explored in the literature.

Here we want to reproduce heat constitutive relations in Landau paradigm. In Landau theory, free energy density under external field can be expanded in power of order parameter. Here we set the magnitude of heat flux $q$ as order parameter and temperature gradient $-\nabla T$ as external field. Free energy density of a thermal system can be expressed as
\begin{equation}
	F=F_0+\frac{a}{2}q^2+\frac{b}{4}q^4+\frac{c}{6}q^6+\dots+ \zeta q\nabla T,\label{freeEnergy}
\end{equation}
where $a,b,c$ are Landau coefficients and $\zeta$ is a coupling constant. Landau-Khalatnikov equation is often used to discuss the behavior of a mateiral near critical point. The Landau-Khalatnikov equation in thermal system is
\begin{equation}
	\Gamma \frac{\partial q}{\partial t} =-\frac{\delta F}{\delta q},\label{LKeq}
\end{equation}
where Khalatnikov kinetic coefficient $\Gamma$ determines the relaxation time $\tau = \frac{\Gamma}{|a|}$ \cite{landauKhalatnikov,LK2021,Size}, $a$ is the first Landau coefficient of free energy density. Traditionally, the order parameter in Landau-Khalatnikov equation is the polarization of feroelectric materials. Here we use thermal flux as order parameter and we argue that Eq.(\ref{LKeq}) is valid for other type of classical transports by taking the order parameter as corresponding conserved current.

To check the validity of Eq.(\ref{LKeq}), we first study the constitutive relations in thermal systems. Take Landau expansion up to second order and suppose the first Landau coefficient is positive, Landau-Khalatnikov equation gives
\begin{equation}
	\tau \frac{\partial q}{\partial t} +q=-\frac{\zeta}{a} \nabla T,\label{MC}
\end{equation}
which is the form of Maxwell-Cattaneo constitutive relation \cite{cattaneo} where heat conductivity is the inverse of first Landau coefficient times coupling constant. If gradient energy term $\frac{\lambda}{2}(\nabla q)^2$ is introduced to the free energy density, Landau-Khalatnikov equation produces Guyer-Krumhansl consititutive relation which is used to study non-local effects in heterogeneous materials \cite{guyer1966solution,ramos2023mathematical}
\begin{equation}
	\tau \frac{\partial q}{\partial t} +q=-\frac{\zeta}{a} \nabla T +\frac{\lambda}{a} \nabla^2 q.
\end{equation}
Notice that current conservation equation $\partial_\mu q^\mu = 0$ cannot be derived from Eq.(\ref{LKeq}), where the 0-th commponent of 4-heat flux $q^\mu=(u,\vec{q})$ is thermal energy. 

Since the exact expression of heat flux is not needed in our method and the heat flux is a conserved current, we propose that our method should be valid for classical transport of different systems by replacing heat flux as Noether current $J$ and negative temperature gradient as the external field. In Supplemental Material we derive the classical Hall conductivity and constitutive relation in London equations of pure material. We claim that the expression of free energy density $F(J)$ is only determined by the defect of a system. It seems  that order parameter in Ginzburg-Landau theory of superconductor should not be the wavefunction, but rather electric current density. And the conserved current of this theory should be the Noether current of the Noether current, which we refer as 2-form Noether current $J^{\{2\}}$.

Here we elucidate the physical interpretion of $J^{\{2\}}$ in U(1) gauge theory. The equations of motion give
\begin{equation}
	\begin{cases}
		\mathrm{d} F=0\\
		\mathrm{d} \star F=\mu_0\star J
	\end{cases},
\end{equation}
where $\rm d$ is exterior derivative and $\star$ is hodge star. By taking exterior derivative on both sides of the second equation we could have
\begin{equation}
	\begin{cases}
		\mathrm{d} J=J^{\{2\}}\\
		\mathrm{d} \star J=0
	\end{cases},
\end{equation}
where $\mathrm{d} \star J=0$ is the conservation law \cite{harlow2021symmetries} and $\mathrm{d} J=J^{\{2\}}$ is the consititutive relation. The expression of $J^{\{2\}}(J)$ should be given by the defects of the theory. Using the conservation condition $\mathrm{d}J^{\{2\}}=0$ and $\mathrm{d} F=0$, we could naturally construct the London gauge $J\propto A + \mathrm{d}\theta$. Using this method one could construct current conservation laws and constitutive relations in higher-form. 
For the general demonstration of arbitrary symmetries, we suggest that Noether's second theorem \cite{noether1918invariante,gieres2023covariant} could be the key, since it is also a bootstrap method and provides constraints on Noether currents.

{\it{Phase transition for perfect thermal conductor}}\quad
Landau-Khalatnikov equation is used to analyze $P$-$E$ hysteresis loop and domain structures in ferroelectrics \cite{LK2021}. By analogy with ferroelectrics, we propose the concept of thermal domains. Inside thermal domain the heat flux parallelly arranged. In Fig.\ref{fig:epsart} , the domain structure can be classified as two categories: unconnencted domain and connencted vortex domain. The vortex domains only occur at perfect conducting regions and heat currrents are localized within domains. 


\begin{figure}[h]
	\subfigure[]
	{
		\begin{minipage}[b]{.4\linewidth}
			\includegraphics{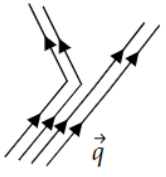}
		\end{minipage}
	}
	\subfigure[]
	{
		\begin{minipage}[b]{.4\linewidth}
			\includegraphics{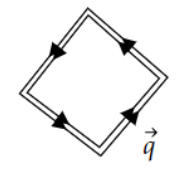}
		\end{minipage}
	}
	\caption{\label{fig:epsart}(a) Unconnected thermal domain. (b) Connected vortex thermal domain.}
\end{figure}

The formation of vortex thermal domains is caused by defect scattering. Here we assume that energy is conserved during scattering process. We take Landau-Khalatnikov equation as Euler-Lagrange equation of an effective field theory. The effective Lagrangian can be derived as
\begin{equation}
	\mathcal{L}_{\rm eff}(q) =-F(q)-\frac{1}{2} \Gamma \dot{q} q,
	\label{EFF}
\end{equation}
where $F$ is free energy density, $\Gamma$ is Khalatnikov kinetic coefficient.
In Supplement Material we show that the effective theory has the similar form of Euclidean complex scalar theory in non-relativistic limit and we could get conserved current version of XY model \cite{Amateur}. We stress that this does not contradict with Coleman-Mermin-Wagner theorem in 2D cases. In this situation the massless Goldstone should also be a current and its momentum $p$ could never reach to zero as long as the fluctuation $G(0,0)=\int\frac{\mathrm{d}^dp}{(2\pi)^d}\frac{1}{p^2}$ does not vanish. Detail discussion is presented in Supplement Material.

We argue that there should be a latent heat in perfect conducting phase transition. As illustrated in Fig.\ref{latent}, in heat releasing process, when temperature gradient is imposed on two side of material, the vortex domain restore the heat, contributing a part of heat capacity. When the temperature approaches to critical temperature, the vortex domains merge into a larger vortex structure and genus of domain structures decreases. When the temperature reaches to critical temperature, the vortex structure might be so large that it connects to unconnected domain or become unstable and release the restored heat into bath. 



\begin{figure}[h]
	\includegraphics[width=1.0\linewidth]{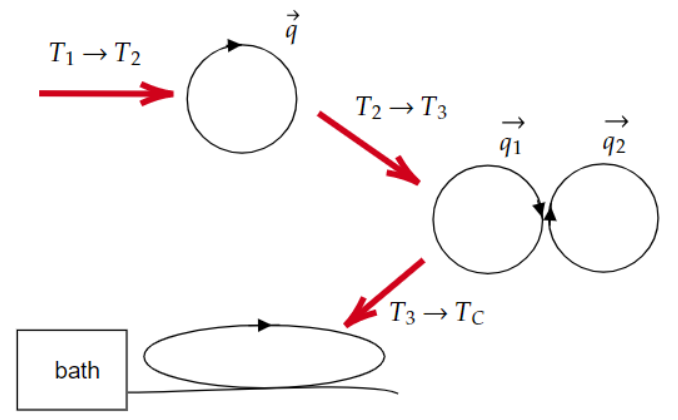}
	\caption{\label{latent} Illustration of how latent heat comes into being. In the first process, vortex domains are formed and contribute to heat capacity. In the second process, when temperature changes, vortices with same direction merge with each other and genus of thermal domains decreases. In the last process, the vortices are so large that at critical point they connect to open-path domains or become unstable, and heat is released to the bath. }
\end{figure}

{\it{Discussion and summary}}\quad
 The heat flux in Eq.(\ref{EFF}) is usually treated as a three dimensional real vector, yet we show that it may well be a complex field. This may indicate that  there might exist anti heat flux. Furthermore, if we impose gauge invariant condition on Landau-Khalatnikov equation, it will lead to the time-dependent Ginzburg-Landau equation \cite{schuller2006time}, suggesting a deeper connection with studies of superconductor \cite{HongLiuLecture, Kapustin_2023,Amateur,BKT}. Spontaneous symmetry breaking can be discussed in this framework.
The thermal version of magnetic field introduced by the gauge field might be used to investigate thermal Hall effect \cite{strohm2005phenomenological,ideue2017giant,hirschberger2015large,lalena2020principles,xu2023}. 

The thermal domain model may provide new understanding of thermal conductivity and latent heat. The vortex domain structure shares conceptual similarities with Anderson localization \cite{anderson1958absence} where electrons are localized in certain areas inside a material. The conductivity decreases such that the material would become an insulator named as Anderson insulator \cite{li2009topological}. It may raise the intriguing possibility of designing a thermal analog of Anderson insulators. 


The ferroelectric anology has been discussed in the context and it is natural to ask whether we could have an antiferroelectric anology of domain structures. It is obvious that antiferroelectric anology cannot be realized in 2D systems since the domain is described by Noether current. In 3D systems, this anology has similar structure as Skyrmion's \cite{skyrme1961non,skyrme1962unified,bhowal2022magnetoelectric}. Altermagnetic anology \cite{krempasky2024altermagnetic} could also be discussed in three-dimensional cases.

The process that objects collapse to a neutron star \cite{burrows1986birth,van1991neutron,aguilera20082d,cumming2017lower} or a black hole \cite{yuan2014hot}
could also be interpreted as phase transitions. If these phase transition have latent heat, then the calculation of latent heat $L=T\Delta S$ could provide the entropy change during the processes. It might provide detailed descriptions of entropy flow and promote the understanding of thermodynamics in cosmology.

When we describe systems with merely Noether current, additional uncertainty will be introduced and the entanglement entropy in our theory should be described using symmetry-resolved entanglement entropy \cite{zhao2021symmetry}. Recently the holographic calculation of symmetry-resolved entanglement entropy has been developed \cite{huang2025symmetry}. It is interesting to see whether our theory could have a holographic description.

In summary, we conjecture that heat consititutive relations can be derived from Landau-Khalatnikov equation when taking heat flux as order parameter. We construct corresponding effective Lagrangian and find that the form of this Lagrangian is very similar to the imaginary time complex scalar field at non-relativistic limit.
We claim that this formalism is also valid for other classical transports and they only differ by the expression of conserved flux $J$. Free energy density $F(J)$ is determined by defects of a specific system. We construct the thermal domain model by analogy with ferroelectrics and qualitatively discuss its phase transition.

\begin{acknowledgments}
	{\it{Acknowledgements}} We acknowledge helpful discussions with Ding Ding and Jiping Huang. 
\end{acknowledgments}

\bibliography{manuscript}

\newpage\hbox{}\thispagestyle{empty}\newpage

\title{Supplemental Material: Emergent Theory for Constitutive Relations in Classical Defect Systems}
\maketitle
\setcounter{equation}{0}
\setcounter{figure}{0}
\setcounter{table}{0}
\setcounter{page}{1}
\makeatletter
\renewcommand{\theequation}{S\arabic{equation}}
\renewcommand{\thefigure}{S\arabic{figure}}
\setcounter{section}{0}
\renewcommand{\thesection}{S-\Roman{section}}

\title{Supplemental Material: Emergent Theory for Constitutive Relations in Classical Defect Systems}

\section{Effective Lagrangian of a thermal system}

The effective Lagrangian of a thermal system is given by
\begin{equation}
	\mathcal{L}_{\rm eff}(q) =-F(q)-\frac{1}{2} \Gamma \dot{q} q,
\end{equation}
where $F$ is free energy density, $\Gamma$ is Landau-Khalatnikov coefficient, q is real heat flux. Landau-Khalatnikov equation can be derived from Euler-Lagrangian equation
\begin{equation}
	\Gamma \frac{\partial q}{\partial t} =-\frac{\delta F}{\delta q}. \label{LKeq}
\end{equation}
We want to find out the contribution of vortex domain to heat conductivity. We introduce energy gradient term $\lambda(\nabla q)^2/2$ into the free energy density to analyze influence of domains. Here we take fourth order of Landau expansion of free energy. 
\begin{equation}
	F=\frac{\lambda}{2}(\nabla q)^2+\frac{a}{2} q^{2} +\frac{b}{4} q^{4} +\zeta q\nabla T,\label{Free}
\end{equation}
and effective Lagrangian is
\begin{equation}
	\mathcal{L}_{\rm eff}=-\frac{\lambda}{2}(\nabla q)^2-\frac{a}{2} q^{2} -\frac{b}{4} q^{4} 
	- \zeta q\nabla T-\frac{1}{2} \Gamma \dot{q} q,\label{EffLag}
\end{equation}
which shares similar form of Euclidean version of complex scalar field at non-relativisic limit. This will be discussed in the next section.

The partition function is defined as $Z[\nabla T]=e^{-\beta F}$. The latent heat can be evaluated as
\begin{equation}
	\langle q\rangle_{T=T_c,\nabla T=0}=-\frac{1}{\beta}\frac{\partial lnZ}{\partial (\nabla T)}\big|_{T=T_c,\nabla T=0}=\zeta q.
\end{equation}
Notice that the expectation value seems to be independent of critical temperature. This implies that the coupling constant $\zeta(T)$ should rely on temperature.

\section{Non-relativistic limit of complex scalar field}

An interacting complex field theory can be written as \cite{Amateur}
\begin{equation}
	\mathcal{L} =\partial ^{\mu } \psi ^{\dagger }( x) \partial _{\mu } \psi ( x) -m^{2} \psi ^{\dagger }( x) \psi ( x) -g\left[ \psi ^{\dagger }( x) \psi ( x)\right]^{2}, \label{Complex}
\end{equation}
with natural unit $c=\hbar=1$. In non-relativistic domain the excitation energy of particles $\epsilon$ is small compared to $mc^2$. The following substitution is used to make a complex scalar field theory non-relativistic
\begin{equation}
	\psi =\frac{1}{\sqrt{2m}} e^{-imt} \Psi.
\end{equation}
Now focusing on the time derivative in Lagrangian. By taking non-relativisic limit we have
\begin{equation}
	\partial _{0} \psi ^{\dagger } \partial _{0} \psi =\frac{1}{2m}\left[ \partial _{0} \Psi ^{\dagger } \partial _{0} \Psi -im\left( \Psi ^{\dagger } \partial _{0} \Psi +\left( \partial _{0} \Psi ^{\dagger }\right) \Psi \right) +m^{2} \Psi ^{\dagger } \Psi \right] .
\end{equation}
The first term is of order 1/m and is negligible. The last term cancels with the mass term in Eq.(\ref{Complex}). $\left( \Psi ^{\dagger } \partial _{0} \Psi +\left( \partial _{0} \Psi ^{\dagger }\right) \Psi \right)$ term can be replaced by $-2\Psi^\dagger\partial_0\Psi$. The final result at non-relativistic limit is
\begin{equation}
	\mathcal{L} =i\Psi ^{\dagger } \partial _{0} \Psi -\frac{1}{2m} \nabla \Psi ^{\dagger } \cdotp \nabla \Psi -\frac{\tilde{g}}{2}\left[ \Psi ^{\dagger } \Psi \right]^{2},
\end{equation}
where $\tilde{g}=g/2m^2$. The choice between mater and antimatter that we simplifying the time derivative term makes the Lagrangian lack of covariance. If We Wick rotate the Lagrangian by $it \to \tau$, we have
\begin{equation}
	\mathcal{L} =-\Psi ^{\dagger } \partial _{\tau} \Psi -\frac{1}{2m} \nabla \Psi ^{\dagger } \cdotp \nabla \Psi -\frac{\tilde{g}}{2}\left[ \Psi ^{\dagger } \Psi \right]^{2}.
\end{equation}
If external potential $\mu\Psi^\dagger(x)\Psi(x)$ and external source $J(x)\Psi^\dagger(x)$ enter into the Lagrangian, it will give
\begin{equation}
	\mathcal{L} =-\Psi ^{\dagger } \partial _{\tau} \Psi -\frac{1}{2m} \nabla \Psi ^{\dagger } \cdotp \nabla \Psi +\mu\Psi ^{\dagger } \Psi -\frac{\tilde{g}}{2}\left[ \Psi ^{\dagger } \Psi \right]^{2} +J\Psi^\dagger,\label{superfluid}
\end{equation}
which shares similar form with Eq.(\ref{EffLag}) with correspondence $\lambda=1/m$, $\mu=-a/2$, $\tilde{g}=b/2$, $\Gamma=2$ and $J(x)=-\zeta\nabla T$. This correspondence tells us that only in perfect conducting situation that the chemical potential $\mu=0$ and that this system is a canonical ensemble. To complexify the heat flux, anti heat flux can be introduced. The physical meaning of anti heat flux can be interpreted as ``bubble flux" in a material. This ``bubble flux" might be caused by vacancy of charge carriers.

\section{XY model}
The complex scalar field in Eq.(\ref{superfluid}) can be represented in polar coordinate form $\Psi(x)=\sqrt{\rho(x)}e^{i\theta(x)}$. Set $J=0$ and Eq.(\ref{superfluid}) becomes
\begin{equation}
	\mathcal{L}=i\rho \partial_0 \theta-\frac{1}{2 m}\left[\frac{1}{4 \rho}(\nabla \rho)^2+\rho(\nabla \theta)^2\right]-\frac{\tilde{g}}{2}(n-\rho)^2,
\end{equation}
where $n=\frac{\mu}{\tilde{g}}$ and total derivative $\partial_\tau \rho/2$ drops out since it has a vanishing contribution to the action. The Lagrangian has a minimal $\rho = n$ and the possible ground states of the theory is $\Psi(x)=\sqrt{\rho(x)}e^{i\theta(x)}=\sqrt{n}e^{i\theta_0}$, where $\theta_0$ is the same at every point in space. Consider the excited state $\sqrt{\rho(x)}=\sqrt{n}+h$ and drop out the total time derivative $i\rho \partial_0 \theta$, the Lagrangian becomes
\begin{equation}
	\mathcal{L}=-\frac{1}{2 m}(\nabla h)^2-2 g n h^2-\left(2i \sqrt{n} \partial_\tau \theta\right) h-\frac{n}{2 m}(\nabla \theta)^2+\ldots.
\end{equation}
It is a Gaussian integral and can be written as
\begin{equation}
	\mathcal{L}=-n \partial_\tau \theta \frac{1}{2 \tilde{g} n-\frac{1}{2 m} \nabla^2} \partial_\tau \theta-\frac{n}{2 m}(\nabla \theta)^2+\ldots.
\end{equation}
At low energy and small momentum cases $1/(2 m) \nabla^2\ll\abs{2gn}$ and the Lagrangian is 
\begin{equation}
	\mathcal{L}=-\frac{1}{2 \tilde{g}}(\frac{\partial \theta}{\partial \tau})^2-\frac{n}{2 m}(\nabla \theta)^2.
\end{equation}
Using rescaling $\tilde{\tau}=\sqrt{\frac{m}{gn}}\tau$, the Lagrangian could be written as
\begin{equation}
	\mathcal{L}=-\frac{n}{2 m}(\partial_\mu \theta)^2,\label{kinetic}
\end{equation}
which is the Lagrangian of XY model when discretized in an orthorhombic grid of points with grid spacings $a$ \cite{chen1997phase,drouin2022kosterlitz}.


Coleman-Mermin-Wagner theorem \cite{Amateur} states that when dimension spontaneous symmetry breaking cannot exist in systems with dimension $d\leq 2$. The propagator of cpmplex scalar field in d-dimensional Euclidean space could be written as $G(x,y)=\int\frac{\mathrm{d}^dp}{(2\pi)^d}\frac{e^{-ip\cdot (x-y)}}{p^2+m^2}$. Since Goldstone particle is massless, the propagator $G(0,0)=\int\frac{\mathrm{d}^dp}{(2\pi)^d}\frac{1}{p^2}$ has a singularity at $p=0$. When $d\leq 2$ the singularity is not integrable and the fluctuations diverge. It is argued that spotaneous symmetry breaking cannot happan for $d\leq 2$. However, here the scalar field in Eq.(\ref{superfluid}) should be some current, so as the Goldstone. If $p=0$, there would be no current and the correlator should be zero. If the current exists, then there must be a lower bound for momentum $p$ and the IR divergence will not occur.

\section{Constitutive relations of 4-current}
In this section we will discuss constitutive relations of 4-current. Here we consider pure material which does not have any disorder. All Landau coefficients in free energy density should vanish and the system is thus a perfect DC conductor. For Maxwell theory, the electromagnetic tensor $F_{\mu\nu}$ is an antisymmetric tensor and $J^\mu J^\nu F_{\mu\nu}=0$. We add this term to free energy density $F[J^\mu]=\frac{\zeta}{2}J^\mu J^\nu F_{\mu\nu}=0$. The x-component of Landau-Khalatnikov equation is
\begin{equation}
	\begin{array}{ l l }
		\Gamma \partial _{t} J_{x} & =-\zeta J^\nu F_{x\nu}\\
		& =-\zeta ( \ J^{0} E_{x} +(\vec{J} \times \vec{B})_{x})\\
		& =-\zeta ( \ J^{0} E_{x} +J_{y} B_{z} -J_{z} B_{y})
	\end{array},\label{Hall}
\end{equation}
where we use the natural units $c=\hbar=1$. If $B_y=0$, then we get Hall conductivity $\sigma_{xy}=-\frac{ J^{0}}{B_{z}}$ \cite{tong2016lectures}. Compare with the result in classical Hall effect we have  $\zeta=1$, $J^0=-ne$, $\Gamma=\frac{m}{e}$ where $m$ is the mass of the carrier, $e$ is the electric charge and $n$ is the density of charge carriers. When $\vec{B}=0$,  the equation becomes the constitutive relation in London equations
\begin{equation}
	\partial _{t} J_{x} =\frac{ne^{2}}{m} E_x.
\end{equation}

The time component of Landau-Khalatnikov equation gives
\begin{equation}
	\begin{array}{ l l }
		\Gamma \partial _{t} J_{0} & =-\zeta\ J^{\nu } F_{0\nu }\\
		& =\zeta(\vec{J} \cdot \vec{E})
	\end{array},
\end{equation}
Using the correspondence of coefficients, it can be written as
\begin{equation}
	m\partial _{t} n=\vec{J} \cdot \vec{E}.
\end{equation}
Now lets consider the Landau-Khalatnikov equation for 4-heat flux $q_\mu=(-u,\vec{q})$. If the thermal system is pure, according to the free energy density given in Eq.(\ref{Free}), the 0th component of Landau-Khalatnikov equation gives
\begin{equation}
	-\Gamma \partial _{t} u=-\zeta \partial _{t} T,
\end{equation}
where $C\equiv\partial u/\partial T=\zeta /\Gamma $ defines the heat capacity.

\section{Ballistic heat conduction}

In semiconductor nanowires, heat conduction becomes non-diffusive for phonons carrying thermal energy experience no diffuse scattering in the small structure \cite{anufriev2021ballistic,tavakoli2018heat,hsiao2013observation}. Here we want to illustrate that ballistic heat conduction should be a perfect thermal conduction. For perfect thermal conductor, according to Eq.(\ref{LKeq}), its heat constitutive relation should be expressed as
\begin{equation}
	\begin{array}{ c c }
		\partial _{t} \mathbf{q} & =-\frac{\zeta }{\Gamma } \nabla T\\
		& =-C\nabla T
	\end{array} , \label{Perfect}
\end{equation}
where $C\equiv\partial u/\partial T=\zeta /\Gamma $ is the heat capacity.


Since the ballistic heat conduction is a non-diffusive classical transport, all Landau coefficients should vanish and heat conductivity should approach to infinity. However, Hsiao et al. \cite{hsiao2013observation} measured the thermal conductivity of SiGe nanowires and found that the heat conductivity $\kappa  \propto L^1$, where L is the length of the system. Here we shall show that this description is compactible to Eq.(\ref{Perfect}).

The phonons experience no diffuse scattering and we can assume that their velocity $v=\Delta x/\Delta t$ is constant. If we use the approximation $\nabla T \simeq \Delta T/\Delta x$, Eq.(\ref{Perfect}) becomes
\begin{equation}
	\Delta q\simeq -\frac{C}{v} \Delta T. \label{Delta}
\end{equation}
If $\zeta$ and $\Gamma$ in Eq.(\ref{Perfect}) is independent of the size of the system, then heat capacity $C$ is also size-independent.
According to the description by Hsiao et al., we could set $\kappa=k\Delta x$ in Fourier constitutive relation
\begin{equation}
	\begin{array}{ c l }
		q & \simeq - \kappa \frac{\Delta T}{\Delta x}\\
		& =-k\Delta x\ \frac{\Delta T}{\Delta x}\\
		& =-k\Delta T
	\end{array} .\label{Ballistic}
\end{equation}
The expression of Eq.(\ref{Ballistic}) and Eq.(\ref{Delta}) only differ by the left hand side. We argue that if the heat flux is measured starting from zero, then $\Delta q =q$, these two expressions are the same.

\end{document}